\documentstyle[12pt]{article}
\def\d{\partial}

\def\a{\alpha}

\def\g{\gamma}
\def\G{\Gamma}

\def\N{\nabla}

\def\de{\delta}
\def\P{\Phi}

\def\e{\varepsilon}

\def\t{\tilde}

\def\is{\equiv}

\def\pmb#1{\setbox0=\hbox{#1}%
\kern.0em\copy0\kern-\wd0
\kern-.04em\copy0\kern-\wd0
\kern.08em\copy0\kern-\wd0
\kern-.04em\raise.0433em\box0 } 	

\def\half{{\textstyle{1 \over 2}}}
\newcommand{\nc}{\newcommand}
\nc{\pek}[1]{\cite{#1}}
\nc{\enr}[1]{(\ref{#1})}
\nc{\kal}[1]{{\cal{#1}}}

\def\bop#1{\setbox0=\hbox{$#1M$}\mkern1.5mu
        \vbox{\hrule height0pt depth.04\ht0
        \hbox{\vrule width.04\ht0 height.9\ht0 \kern.9\ht0
        \vrule width.04\ht0}\hrule height.04\ht0}\mkern1.5mu}

\begin{document}

\newcommand{\inv}[1]{{#1}^{-1}} 

\renewcommand{\theequation}{\thesection.\arabic{equation}}
\newcommand{\beq}{\begin{equation}}
\newcommand{\eeq}[1]{\label{#1}\end{equation}}
\newcommand{\ber}{\begin{eqnarray}}
\newcommand{\eer}[1]{\label{#1}\end{eqnarray}}
\begin{center}

                                \hfill    hep-th/9910159\\

\vskip .3in \noindent

\vskip .1in

{\large \bf {Limits of the $D$-brane action }}
\vskip .2in

{\bf Ulf Lindstr\"om}$^a$\footnote{e-mail address: ul@physto.se}, {\bf Maxim
Zabzine
$^a$\footnote{e-mail address: zabzin@physto.se} and {\bf Aleksandr
Zheltukhin }$^{a,b}$\footnote{e-mail address: aaz@physto.se} \\

\vskip .15in

\vskip .15in

\vskip .15in
$\mbox{}^{a)}$ {\em  Institute of Theoretical Physics,
University of Stockholm \\
Box 6730,
S-113 85 Stockholm SWEDEN}\\}
\bigskip

$\mbox{}^{b)}$ {\em Kharkov Institute of Physics and Technology,\\ 
310108, Kharkov, Ukraine\\}
\vskip .15in

\vskip .1in
\end{center}
\vskip .4in
\begin{center} {\bf ABSTRACT } 
\end{center}
\begin{quotation}\noindent For background geometries whose metric contain
a scale $\g$ we reformulate the Born-Infeld
$D$-brane action in terms of $\e \equiv \g /(2\pi \alpha ')$. This may be
taken as a starting point for various perturbative treatments of the theory.
We study two limits that arise at zeroth order of such perturbations. In
the first limit, that corresponds to the $g_s\to\infty$ with $\e$ fix, we
find a "string parton" picture, also in the presense of some background
$RR$-fields. In the second limit, $\e\to 0$, we find a topological model.

\end{quotation}
\vfill
\eject

\section{Introduction}

In this brief letter we study two limits of the Born-Infeld $D$-brane
action. We base the treatment on an expansion in a rescaled tension $\e$,
and include a full background.

We first take a new look at a "strong
coupling" limit ($g_s \to \infty$). This is
intended as a supplement to the discussion in \cite{ulrvu} and
\cite{hgul} where the $D$-brane world volume was shown to be
foliated by string world sheets in this limit.  The result is a corroboration
of the foliation picture plus the modified
equations involving the background $n$-forms. In particular, we give an
example showing that the allowed "string partons" are either infinite
(which might correspond to $D1$-branes), closed or open strings ending on
other
$D$-branes (sources of the
$RR$-fields ).  We also show that the string theory must be of type
$IIB$.

The second limit we present is $\e \to 0$. Here we give an $\e$ expansion to
first order and discover that the leading term is topological, i.e.
independent of the metric. For the case when the dilaton and the $0$-form
are constant it is proportional to the Chern class of the $U(1)$-field.

We suggest that both these limits might serve as suitable starting points
for investigating $D$-branes order by order in a parameter expansion. To do
so one needs to specify the background, however, and we leave this as a
topic for future studies.

\section{Background}

The long wavelength limit of $D$-brane dynamics is described
by the Born-Infeld action
\beq
S_{BI}(T_{p})=T_{p}\int d^{p+1}\xi e^{-\P '}\sqrt{-\det(
       \g_{ij}+\kal{F}_{ij})},
\eeq{binf}
where $\g _{ij}\equiv \d _iX^\mu \d_jX^\nu G_{\mu\nu}(X)$ is the
 metric on the world-sheet induced from a background metric $G_{\mu\nu}$ and
${\cal F}_{ij}$ is a world-volume two-form,
\beq
{\cal F}_{ij}\equiv 2\pi\a '\d_{[i}A_{j]}+B_{ij} .
\eeq{fdef}
Here $B_{ij}\equiv\d_iX^\mu\d_jX^\nu B_{\mu\nu}$ is the pull-back of the
(dimensionless) background Kalb-Ramond field. The $p$-brane tension $T_p$ is related
to the fundamental string tension $T\equiv (2\pi \a ')^{-1}$
and the string coupling constant $g_s\equiv e^{\P_\infty}$ by
\beq
T_p\equiv {1 \over {(2\pi )^pg_s(\sqrt{\a '})^{p+1}}},
\eeq{Tp}
with $\P_\infty$ being the dilaton expectation value and $\P
'\equiv\P -\P_\infty$. 

We shall be interested in the case when the background metric
$G_{\mu\nu}$ contains a parameter $\g$ of dimension $length^2$.
As described in \cite{zhel} this can then be used to introduce a
perturbative scheme in the rescaled  string tension\footnote{Calling $\e$
the tension is really a misnomer. It is a dimensionless entity and only
related to the fundamental string tension via a rescaling.  We hope that
this will not cause any confusion.}
$\e\equiv{\g\over{2\pi\a '}}$. The parameter $\g$ may,
e.g., be (the square of) the radius of (anti) deSitter
space. In general $\g$ will be choosen from the parameters
in the string moduli space, and this choice will not be unique.

Rescaling the fields according to
\beq 
\t\xi^i = \g^{-\half}\xi^i,\quad\t X^\mu =\g^{-\half}X^\mu,\quad
\t A_\mu = \g^\half A_\mu,
\eeq{resc}
and including the additional coupling to the various (dimensionless)
antisymmetric forms $C_{\mu_1...\mu_k}$, the $D$-brane action in terms of
dimensionless fields reads
\beq
S=\t T_p\left[\int d^{p+1}\t\xi e^{-\P '}\sqrt{-\det(
       \t\g_{ij}+\t{\cal F}_{ij})} +g_s\int\t C\wedge
e^{\t \kal{F}}\right],
\eeq{react}
where
\beq
\t T_p \is 
{
{\e^{{p+1}\over 2}}
\over
{g_s(2\pi)^{{p-1}\over 2}}},\,\,\,\,\,\,\,\,\,\,\,\,\,\,\,
 \t{\cal F}_{ij} = \frac{1}{\e} \t\d_{[i}\t{A}_{j]}+\t{B}_{ij} .
\eeq{tildeT}

The action \ref{react} is written in the string frame. The
dilaton (coupling constant) dependence of the brane-action is frame
dependent. We choose to consider the string frame since here the mass of
fundamental strings is $\kal{O}(1)$ in the string coupling and thus
unaffected by our strong coupling limit. Other limits have been considered,
however \cite{berto}.

\section{A strong coupling limit}

In this section we will study a limit of the $D$-brane action
when $g_s\to \infty$ in the context of Type $IIB$ string theory.

In general, to investigate the strong coupling limit of string-
or
$M$ theory one would like to consider the limit of the action
(\ref{react}) for
large $g_s$ in conjunction with a transformation of
the background \cite{Chris}. We will address the simpler problem
of taking the limit $g_s\to \infty$, or more precisely
${
{\e^{{p+1}\over 2}}\over g_s} <<1,\quad \e =const$, in a fixed
background. 

The restriction to Type $IIB$
theory is now evident from the following: In Type
$IIA$ string theory, because of its $11D$
$M$-theory origin, the string coupling $g_s$ is related to the
radius
$R_{11}$ of the compactified dimension through  $g_s^{2\over
3}\sqrt{\a '}= R_{11}$. Hence $g_s^{4\over
3}=\e $, (identifying $\g =R_{11}^2/2\pi$ in the definition of $\e$ above),
and it is inconsistent to let
$g_s\to
\infty$ keeping
$\e$ fixed.

Using the method employed in \cite{ulrvu} and \cite{hgul}, we
find the desired limit to be (dropping tildes),
\beq
S=\half\int d^{p+1}\xi\left[V^iW^j(
       \g_{ij}+{\cal F}_{ij})\right]+\e^{{p+1}\over 2}\int
C\wedge e^{ \kal{F}},
\eeq{vact}
where $V^i(\xi)$ and $W^j(\xi)$ are world volume vector
densities. 
This may be considered as the zeroth order term in an expansion
of the action (\ref{react}) in powers of $g_s^{-1}$. We will
study it in its own right, though.

The $V^i$ and $W^j$-field equations derived from (\ref{vact})
read
\ber
W^j(\g_{ij}+{\cal F}_{ij})&=&0 \cr
V^i(\g_{ij}+{\cal F}_{ij})&=&0,
\eer{feqs}
  and 
 the integrability condition\footnote{We shall not be interested in the trivial
 solution $V^i=W^i=0$. This corresponds to a theory with an WZ action given by the 
 second term on the r.h.s. of (\ref{vact}).}
 for these equations  is
\beq
\det(\g_{ij}+{\cal F}_{ij})
\is\det(\g_{ij}+B_{ij}+\e^{-1}F_{ij})=0. 
\eeq{eom1}

To gain a first qualitative understanding of this relation, we
set $B=0$. In this case it is
equivalent to
\beq
\det(\de^i_j+\e^{-1}F^i_{\ j})=0. 
\eeq{eom2}
In analogy to the usual treatment of the Born-Infeld action one
might be tempted to employ a weak gradient expansion \cite{taylor}
and expand (\ref{eom2}) in a power series in $|\d A|<<1$.
However, as is easily seen, this implies that $\e^2 \propto
trF^2<<1$, which is not true in general. E.g., when the background is
Minkowski space we have $\e \propto \infty$\footnote{This results if we think of Minkowski
 space as the $R\rightarrow \infty$ limit of de Sitter space. In an alternative
 scheme discussed in \cite{zhel} $\e$ is of order $1$ which also violates our
 inequality.}.
We thus learn that the  weak gradient expansion is not in general viable.

\subsection{Solving the VW-equations}

Although the equation (\ref{eom1}) says that the combined
matrix $\g +{\cal F}$ is degenerate, this is not so for the
induced metric $\g$ itself. We may thus introduce a vielbein
$e_a^i$ and its inverse $e^a_i$ on the world-volume according to
\beq
e^a_ie^b_j\eta_{ab}=\g_{ij} \quad a,b=0,...,p,
\eeq{vielb}
where $\eta_{ab}$ is the $p+1$-dimensional Minkowski metric. In \cite{ulrvu}
it was found that the first part of the action (\ref{vact}) has a
two-dimensional Lorentz structure. This may be displayed by a redefinition
of $V^i$ and $W^i$.   With
$e^{-1}$ being the determinant of the inverse vielbein, we use the
$SO(p,1)$-tangent space group to write\footnote{We introduce
the tilde notation for later convenience.}
\ber
V^i&=&(e^i_0+e^i_1)\sqrt{e^{-1}}\is (\t e^i_0+\t e^i_1),\cr
W^i&=&(e^i_0-e^i_1)\sqrt{e^{-1}}\is (\t e^i_0-\t e^i_1),
\eer{vwe}
thus breaking the tangent-space group down to
$SO(1,1)\times SO(p-2)$. In this gauge, (\ref{feqs}) reads
\beq
{\cal F}_{ij}e^j_0=\g_{ij}e^j_1, \quad
{\cal F}_{ij}e^j_1=\g_{ij}e^j_0,
\eeq{feqn}
or, equivalently,
\beq
\left({\cal F}^2\right)^i_{\ j}e^j_{0,1}=e^i_{0,1}
\eeq{}
with the integrability condition
\beq
\det\left(\de^i_j-\left({\cal F}^2\right)^i_{\ j}\right)=0.
\eeq{int2}
Expanding the two-form ${\cal F}$ in the bivector basis
\beq
{\cal F}= \half\kal{F}_{ij}d\xi^i\wedge
d\xi^j=\half\kal{F}_{ab}e^a\wedge e^b,
\eeq{}
(\ref{feqn}) is solved by
\beq
{\cal F}=\kal{F}^\|+\kal{F}^\bot =2e^0\wedge e^1+{\cal
F}_{AB}e^A\wedge e^B,
\quad A,B=2,...,p,
\eeq{fsln}
or, equivalently, 

\ber
\N_{[0}A_{1]}&=&\e(1-B_{01}), \quad
\N_{[0}A_{A]}=-\e B_{0A},\cr
\N_{[1}A_{A]}&=&-\e B_{1A},
\quad {\cal F}_{AB}=\e^{-1}\N_{[A}A_{B]}+B_{AB},
\eer{Acnd}
where $\N \is\d +\omega\cdot M$ is the covariant derivative,
$\omega$ being the spin-connection and $M$ the Lorentz-group
generator. 

The integrability condition (\ref{int2}) is of course
identically satisfied by (\ref{fsln}). 

We thus learn from the $V$ and $W$
equations that the 2-form ${\cal F}$ splits into one part which
lies in the $2D$ tangent space spanned by $e^0$ and $e^1$, and
one part which lies entirely in the orthogonal part spanned by
the $e^A$'s. As seen from (\ref{Acnd}), when $B=0$, this
implies that the $D$-brane electric field is constant and lies
in the $e_1$ direction.

\subsection{The A-equations}

The $A$ field equations that follow from (\ref{vact}) are
\ber
\d_i\left(V^{[i}W^{j]}\right)
=\kal{J}^j
\eer{afeq}
where $\kal{J}^i$ represents the contribution from the
background $RR$ gauge fields and is given by
\ber
\kal{J}^i\is
-\e^{(p+1)/2}\sum_{s=1}^{(p+1)/2}
{1\over{(s-1)!2^{s-1}}}
{}^*H^{il_1k_1...l_{s-1}k_{s-1}}\kal{F}_{l_1k_1}...\kal{F}_{l_{s-1}k_{s-1}},
\eer{cur}
where ${}^*H^{l_1...l_{p-n}}$ is the dual of the pull-back of the field
strength of the
$C^{(n)}$ form.

In \cite{ulrvu} and \cite{hgul} the structure of the theory was
shown to be simplified in the following diffeomorphism gauge:
\beq
\d_iV^i=\d_iW^i=0.
\eeq{dgau}
In that gauge (\ref{afeq}) may be rewritten as
\beq
\left[V,W\right]^i=\kal{J}^i
\eeq{lbra}
where $[.,.]$ denotes the Lie-bracket.

The equations (\ref{afeq}) are highly non-linear. To
proceed further we need to make some simplifying assumptions,
and we first consider the case when the background $RR$ gauge
fields are zero $(C^{(n)}=0)$. In that case (\ref{dgau}) says
that $[V^i\d_i,W^j\d_j]=0$ which means that those directions
may be choosen as coordinate-directions
\cite{ulrvu}. In view of (\ref{vwe}) we choose to use the $\t e$
directions instead. 
Explicitly, we thus use a diffeomorphism
gauge where\footnote{We may also use the residual symmetry
preserving \enr{dgau} to set $e=1$.}
\beq
 e^i_a = \left\{e^{1\over 2}\de^i_0,e^{1\over
2}\de^i_1, e^i_A\right\},
\eeq{dgag}
which has the inverse
\ber
&& e_i^{\ a} = \left\{e^{-{1\over 2}}\de^a_0,e^{-{1\over
2}}\de^a_1, e^a_M\right\},\quad
M=2,...,p\cr
&&e_M^{\ A}e_A^{\ N}=\de_M^N,\quad
e_A^{\ M}e_M^{\ B}=\de_A^B, \quad e_M^{\ 0,1}=-e^{-{1\over
2}}e_M^{\ A}e_A^{\ 0,1}.
\eer{idgag}
Again setting $B=0$ and returning to equation (\ref{feqn}), we
find
\beq
F^i_{\ 0}=\e \de^i_1,\quad F^i_{\ 1}=\e \de^i_0.
\eeq{fere}
We see that (the mixed components of) the $U(1)$ curvature now
splits into one part in the $2D$-plane spanned by $\xi^0$ and
$\xi^1$, and one part in the complementary directions spanned
by the $\xi^M$'s. Alternatively, the covariant
components are, (setting $e=1$,)
\beq
F_{i_ 0}= \e\d_i X^\mu \d_1X^\nu G_{\mu\nu}~,\quad F_{i1}=\e\d_i
X^\mu
\d_0{X}^\nu G_{\mu\nu} ~,
\eeq{fere2}
with $F_{AB}$ arbitrary. This is the $\kal{F}$-solution. This
gauge also leads the following expressions that result from
(\ref{vielb}):
\ber
&&\dot{X}^\mu\dot{X}^\nu G_{\mu\nu}+{X'}^\mu {X'}^\nu
G_{\mu\nu}\is\g_{00}+\g_{11}=\eta_{00}+\eta_{11}=0\cr
&&\dot{X}^\mu{X'}^\nu G_{\mu\nu}\is\g_{01}=\eta_{01}=0
\eer{vira}
i.e., the Virasoro constraints, (parametrized by $\xi^i, i\ne
0,1$), in conformal gauge.

\eject
\subsection{The $X^\mu$-equations}

The $X^\mu$ field equations that follow from (\ref{vact}) are
\ber
&&\d_i\left((V^{(i}W^{j)}G_{\mu\nu}+V^{[i}W^{j]}B_{\mu\nu})
\d_jX^\nu\right)\cr
&-&V^iW^j\d_iX^\rho\d_jX^\nu (G_{\rho\nu}+B_{\rho\nu}),_\mu\cr
&=&I_\mu +K_\mu,
\eer{xeqn}
where $I_\mu$ contains the contribution from the $RR$ gauge fields
\ber
I_\mu \equiv
&\e^{(p+1)/2}&\sum_{s=0}^{(p+1)/2}{1\over{(p+1-2s)!2^ss!}}
\d_{i_1} X^{\mu_1}\d_{i_2}
X^{\mu_2}...\d_{i_{(p+1-2s)}}X^{\mu_{(p+1-2s)}}\cr
&&~  \cr
&&\epsilon^{i_1...i_{(p+1-2s)j_1k_1...j_sk_s}}H_{\mu \mu_1...\mu_{(p+1-s)}}
\kal{F}_{j_1k_1}...\kal{F}_{j_sk_s},
\eer{xxeqn}
and $K_\mu$ contains the 
$B$-field strength. Setting the $C^{n}$'s to zero and using the
gauges (\ref{vwe}), (\ref{dgag}), the equation (\ref{xeqn}) becomes
\cite{hgul}
\beq
\d^2_0X^\mu-\d^2_1X^\mu+\G_{\nu\rho}^{\ \
\mu}\d_0X^\nu\d_0X^\rho -\G_{\nu\rho}^{\ \
\mu}\d_1X^\nu\d_1X^\rho=0,
\eeq{}
i.e., the equation of motion of a string in the
$01$-coordinates in conformal gauge. Since we also have the Virasoro
constraints (\ref{vira}), we see that the stringy picture is complete.

\subsection{Comments}

The interpretation of the strong coupling limit that arises from
the above considerations when $C^{(n)}=0$ is as follows: The
world-volume is foliated by world-sheets of (generally $(p,q)$ charged)
strings. These strings are either closed or infinite.  The form
${\cal F}$ is likewise split into one component in the world-sheet
direction and the rest in the complementary directions. When
$B=0$ this splitting applies to the $U(1)$ field-strength
itself: The component in the world-sheet is then determined by
the string tension $\e$ and those in the complementary
directions are not determined by the field equations, in
agreement with the role of the A-field as a Lagrange multiplier
(in this limit).

To gain some further insight into the "string parton" picture one may argue
as follows: If instead of the gauge choice (\ref{dgag}) we only make a
partial choice $\t e_0^i=\delta ^i_0$, we find from the
$A$-equations (\ref{afeq}) that (in the absence of $RR$-fields),
\beq
\d_0 \t e^m_1= \d_m \t e^m_1=0, \quad m\equiv1,...,p.
\eeq{Oaeq}
Thus  $\t e^m_1$ is independent of $\xi^0$ and divergence-free in the
spatial indices $m$. This means that $\t e^m_1$ has no sources and that its
field lines are either closed or go off to infinity. Since the
relations (\ref{dgau}) still hold in this gauge, we retain the string
interpretation with the $\t e^m_1$-direction representing the spatial string
direction. We thus conclude that the strings are either closed or infinite.
The strings going off to infinity might possibly have an interpretation as
$D1$-branes (corresponding to branes within branes).

When $C^{(n)}\ne 0$, the situation will in general be more
complicated. The one exception is when only $C^{(p+1)}\ne 0$,
since this field does not enter in (\ref{lbra}). The above
analysis thus goes through, but now there is a
$C^{(p+1)}$-dependent source on the right-hand side of
(\ref{xeqn}). When $C^{(n)}\ne 0, n<p+1$, the equations become
very hard to analyze, in general. There are special cases, however, where an
analysis similiar to the one above is possible. We now discuss one such
example.

Staying in the gauge $\t e_0^i=\delta ^i_0$ and making the further gauge
choice $\d_0\t e_1^0=0$,  we find from the
$A$-equations (\ref{afeq}) that
\beq
[\t e_0,\t e_1]^0=0, \quad [\t e_0,\t e_1]^m=\kal{J}^m ~.
\eeq{jaeq}
Hence only the spatial components of the currernt (\ref{cur}) , $\kal{J}^m$,
enter the commutation relation, not $\kal{J}^0$. Clearly, in backgrounds
where $\kal{J}^m=0$ we may again use the coordinates (\ref{dgag}) (with $e=1$
for simplicity), and the string interpretation is again viable.

To gain an understanding of the circumstances that might yield such a
result, let us consider the current (\ref{cur}) for the special case 
when only $C^{(p+1)}$ and $C^{(p-1)}\ne 0$. The first of these will not
affect (\ref{jaeq}), as discussed above. The statement that
$\kal{J}^m=0$ is then tantamount to ${}^*H^m=0$. A sufficient condition for
this to be true is that 
\beq
\d_0 X^{\mu_1}H_{\mu_1...\mu_p}=0.
\eeq{}
This last
condition has a rather clear physical meaning: Since we may now go to
coordintes (\ref{dgag}) where the string interpretation holds, we can restate
it as $P^{\mu_1}H_{\mu_1...\mu_p}=0$, i.e., the fieldstrength $H^{(p)}$ is
orthogonal to the momentum-density $P$ of the "parton" string. Since
furthermore our gauge choice $\d_0e^0_1=0$ implies that $\d_m
e^m_1=\kal{J}^0$, the argument below (\ref{Oaeq}) extends to say that in
this case the "string partons" are either infinite strings, closed strings
or open strings that end on the
$D$-branes that are sources for
$C^{(p+1)}$ and $C^{(p-1)}$. The last case requires that these sources
intersect with the world-volume of the $D$-brane under discussion
\cite{doug}.

\section{The $\e \to 0$ limit.}

In this section we turn to the $\e \to 0$ limit of the action
(\ref{react}). This may be interpreted as either a small
tension or a large curvature limit. The latter interpretation is of
particular interest for the AdS/CFT correspondence \cite{malda}, where the
limit away from weak gravitation corresponds to a limit away from
large $N$ in the boundary theory. 
In any case $\e \to 0$ represents a high-energy limit where
higher derivative terms, i.e., derivatives of the
$U(1)$ field strength, are expected to become important. Since we are
studying a low energy effective action this limit thus seems irrelevant at
first sight. Here we take the attitude, however, that there may be features
of the high energy theory that are of topologic and algebraic nature and may
be captured in such a limit. Alternatively, since we do not prove this, one
may also study the limit of the model considered in its own right (i.e.
independent of its relation to string theory).

We again consider the Type$IIB$ theory, which means that $p$
is restricted to be odd. We are interested in the first few
terms in an $\e$- expansion of the action. To that end we
rewrite the Born-Infeld part of (\ref{react}) as
\ber
S&=&g_s^{-1}\int d\xi^1\wedge ...\wedge d\xi^{p+1}e^{-\P}
\left[\det \g
\det \{(\e B +F)_\a^{\ \beta}\}\right.\cr
&&\left.\det\{(\de +
\e (\e B+F)^{-1})_\beta^{\ \g}\}\right]^{\half}.
\eer{}
Making use of the expansion\footnote{This corresponds to the
expansion of the determinant of a $2l\times 2l$ skew matrix as
the square of the Pfaffian.}
\beq
d\xi^1\wedge ...\wedge d\xi^{p+1}\sqrt{\det \{(\e B
+F)\}}={1\over{l!}}(\e B+F)\wedge...\wedge (\e B+F),
\eeq{}
where $2l=p+1$, we expand the full action (\ref{react}) as
follows
\eject
\ber 
S&=&{g_s^{-1}\over{l!}}\int 
\left\{(e^{-\P}+g_sC^{(0)})F\wedge ...\wedge F\right. \cr
&&\left. +l\e((e^{-\P}+g_sC^{(0)})B+g_sC^{(2)})F\wedge ...\wedge
F+O(\e^3)\right\}.
\eer{eexac}
We see that the leading term corresponds to a coupling to a
$D$-instanton. With $q(X) \equiv e^{-\phi}+g_sC^{(0)}$ it has
the form 
\beq
\int
qF\wedge ...\wedge F~, 
\eeq{topo}
which displays its topological character: When $q=const$ it is
proportional to the Chern class of the $U(1)$ field, and it is
always independent of a metric.

The subleading is a (generalized)
Chern-Simons type term. Further we note that the form of the
action (\ref{eexac}) preserves $SL(2,Z)$-invariance. In fact,
had we only considered the Born-Infeld action in this limit
and then imposed $SL(2,Z)$-invariance, we would have
discovered the $C^{(0)}$ and $C^{(2)}$ couplings.\\

\section{Discussion}

We have studied two limits the Born-Infeld action for $D$-branes
in a given fixed background. As a basis for this we used an
action with rescaled fields and tension ($\e$). The strong
coupling limit we took to be defined by $\e^{{p+1}\over 2}/g_s<<1$
with $\e$ fixed and the high energy limit we studied was $\e
\to 0$. 

In the first limit we were able to solve the
$V^i,W^i,A_i$ and $X^\mu$ equations for $C^{(n)}=0, n=0,...,p$
and recover the picture of the world volume as foliated by
strings from \cite{ulrvu}. Restrictions on the geometry led to the 
possible constituent strings going off to infinity ($D1$-branes) or being
closed.The
general case proved to be considerably more complicated due to the
presence of the
current built from the $C^{(n)}$'s, although we saw that the
"string parton" picture survived in a particular example. In that example we
found the additional option of the constituent strings being open but ending
on the
$D$-brane sources of the
$C^{(n)}$ background fields.

In the second limit, we found an $\e$-expansion whose leading
term is topological. We also noted that $SL(2,Z)$-invariance
predicts the form of the $C^{(0)}$ and $C^{(2)}$ couplings in
the next to leading term. We also noted that interpreting this limit as
taking the radius
$R\to 0$ in the AdS picture, this limit should probe physics away from the
large $N$ limit.

To proceed further with both these expansions one would need to
specify the background. In fact, solving the $A_i$ and 
$X^\mu$ equations in the second limit order by order in $\e$
requires knowledge of the $\a'$-dependence of the background.
This is a possible direction for future research. It would also be
interesing to compare our results to other ways of extracting string
dynamics from Born Infeld theory, e.g. the Born-Infeld strings of
\cite{calmal} 
\bigskip

\bigskip

{\bf Acknowledgement}: The research of UL was supported in part by NFR grant
No.\ F-AA/FU 04038-312. AZ was supported in part by a grant from the Royal
Swedish Academy of Sciences. We are grateful to Rikard von Unge for comments
on the manuscript and to S. Thiesen for discussions.

\end{document}